\documentstyle{article}
\input epsf
\begin{document}
\newcommand{\Ts}{T_{\rm sys}}
\newcommand{\Cel}{$^{\circ}$C}
\newcommand{\Fsp}{\vspace{0mm}}

\begin{center}
{\bf Galactic Free-Free and H$\alpha$ Emission}  \\
{\sc GEORGE F. SMOOT} \\
LBNL, SSL, Physics Department \\
University of California \\
Berkeley, CA 94720 \\
e-mail: Smoot@cosmos.lbl.gov \\
\vspace{2mm}
\end{center}


{\bf Abstract:} This document provides a brief summary estimate
of Galactic free-free emission and H$\alpha$ emission and their relationship. 
Particular emphasis is placed on estimating the potential free-free emission 
in the region of significant confusion for CMB anisotropy measurements.
Existing x-ray, ultraviolet and H$\alpha$ emission provide limits on 
the radio free-free emission and vice versa.
These limits are generally somewhat smaller than the observed
``free-free'' (signal $\propto \nu^{-2.15}$) microwave signal.
If these preliminary results true, then some previously neglected source 
may be present.
Physics argues that H$\alpha$ emission is still the best
tracer for Galactic free-free emission and thus a tool for
diagnosing if there is a previously neglected source.

\section{INTRODUCTION}

\begin{figure}[h]
\epsfxsize=5.9truein
\epsfbox{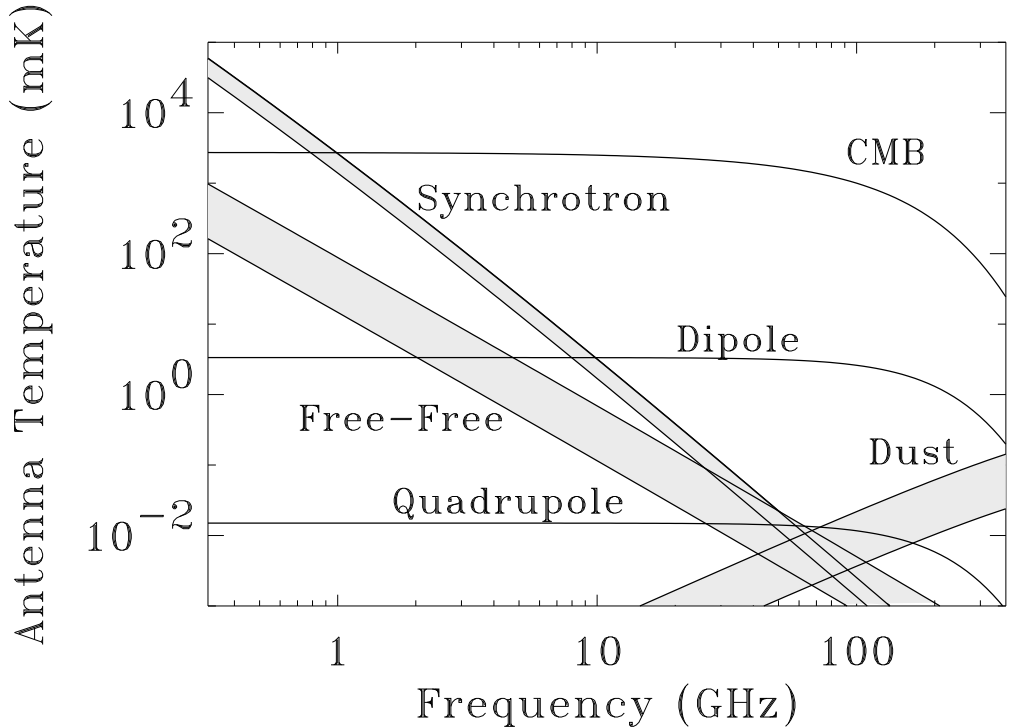}
\caption{
Graph showing frequency dependence and approximate relative strength
of Galactic synchrotron, free-free, and dust emssion
as well as that of the cosmic microwave background and its features.
}
\label{FigGal}
\end{figure}

Free-free emission is the least well known of the three diffuse Galactic
emissions which dominate the mm and cm wavelength sky.
Figure \ref{FigGal} shows versus frequency 
the approximate relative intensity of
the Galactic synchrotron, free-free, and dust emission
in relation to the cosmic microwave background (CMB).

\section{Free-Free Emission}
Radio-frequency free-free emission arises from the interaction 
of free electrons with ions, and consists of thermal bremsstrahlung radiation. 
Free-free emission is the bremsstrahlung (braking radiation)
that occurs when a fast charged particle (in this astrophysical case
thermally hot electrons) is accelerated in an encounter with
an atom, molecule, or ion.
Free-free emission is so named since the electron starts in a free
(unbound) state and ends in an unbound state as opposed to being
captured into a bound state (free-bound emission) or making a transition
from one bound level to another (bound-bound emission).
When the electron enters the Coulomb field, it is accelerated and
emits radiation in a readily calculable manner.
In an astrophysical plasma, such as those that exist in our Galaxy
one must integrate over the total distribution of electrons
and ions to obtain an expression for the volume or line of sight emissivity.
The velocity averaged Gaunt factor $< g_{eff} >$ takes into account
the distributions and the quantum mechanical cutoff in scattering emission.

The volume absorptivity/emissivity of a plasma for electron-ion
bremsstrahlung is \cite{Oster}, \cite{Rybicki}
\begin{eqnarray}
\alpha_\nu &=& {4 e^2 \over 3 m_e h c} ({2 \pi \over 3 k m_e})^2 N_e N_i Z^2
T_e^{-1/2} \nu^{-3} (1 - e^{-h \nu /k T_e} ) < g_{ff}> \\
&=& 3.7 \times 10^8 ~ Z^2 N_e N_i T_e^{-1/2} \nu^{-3} (1 - e^{-h \nu /k T_e} ) 
< g_{ff}> \\
&\approx& {4 e^2 \over 3 m_e h c} ({2 \pi \over 3 k m_e})^2 N_e N_i Z^2
T_e^{-3/2} \nu^{-2} < g_{ff}>  \\
& = & 0.018 ~ Z^2 N_e N_i T_e^{-3/2} \nu^{-2} < g_{ff}>
\end{eqnarray}
where the approximation is for $h \nu << k T_e$ and $< g_{ff}>$
is the velocity averaged Gaunt factor.
Integration along the line of sight gives the formula 
for the optical depth $\tau_\nu$ to free-free as
\begin{eqnarray}
\tau_\nu 
&=& 0.018 ~ T_e^{-3/2} \nu^{-2} \int N_e N_i d l < g_{ff}> 
= 0.018 ~ T_e^{-3/2} \nu^{-2} EM < g_{ff}> \\
&=& 0.0543 ~ T_e^{-1.5} ( {\nu \over 1~GHz })^{-2} ( {EM \over cm^{-6} pc} ) 
< g_{ff}> \\
&\approx& 0.08235 ~ T_e^{-1.35} ( {\nu \over 1~GHz })^{-2.1} 
( {EM \over cm^{-6} pc} ) 
\end{eqnarray}
where the emission measure $EM$ is defined as:
$EM  \equiv \int N_e N_i d l$.
The brightness spectrum has a $-2.1 \pm 0.03$ spectral index, 
with weak dependency on the temperature and density of the interstellar plasma 
and the observing frequency. 
The Gaunt factor accounts for this small dependency.

This formula (Eqn. 7) is based on the assumptions 
that the interstellar plasma is electrically neutral, 
that the temperature of the electrons along the line of sight 
is roughly constant, 
that the electron temperature is greater than 20K, 
the frequency is smaller than 100 GHz, 
and that the Gaunt factor can be expressed as a product of powers 
of the frequency and the electron temperature. 
\cite{Rohlfs},\cite{Rybicki}. 

An estimate of the brightness temperature for free-free 
emission can be found from the radiative transport equations
\begin{eqnarray}
T_b^{ff} &=& T_e (1 - e^{-\tau_\nu} ) \approx  \tau_\nu T_e \\
&\approx& 5.43 \mu K ~ ( {10~GHz \over \nu} )^2 ({ 10^4 ~K \over T_e})^{1/2} 
  ( {EM  \over cm^{-6} pc} ) < g_{ff} > \\
& \sim & 26 \mu K ~ ( {10~GHz \over \nu} )^{2.1} ({ 10^4 ~K \over T_e})^{0.35}
( {EM  \over cm^{-6} pc} )
\end{eqnarray}
An approximation for the velocity-averaged Gaunt factor $< g(\nu,T_e) >$ 
is \cite{Spitzer78}: 
\begin{equation}
< g_{ff} > \approx 4.69 \times 
[ 1 + 0.176 ln(T_e/10^4~K) - 0.118 ln(\nu/10~GHz) ] 
\end{equation}
detail tables and formula can be found many places \cite{Rybicki}.

\section{Free-Free Signal}

Free-free emission is the least well known of the Galactic emissions.
It is not easily indentified at radio frequencies,
except near the Galactic plane.
At higher latitudes free-free emission must be separated 
from synchrotron emission 
by virtue of their differing spectral indices.
At frequencies less than about 10 GHz synchrotron emission dominates
at intermediate and high latitudes.
At higher frequencies where free-free emission might be expected to exceed
the synchrotron component,
the signals are weak and survey zero levels are indeterminate.

\section{H$\alpha$ Emission}
At present most of the information currently available 
about the source of free-free radiation at intermediate and high latitudes
comes from H$\alpha$ surveys.

\subsection{Relation between H$\alpha$ \& Free-free Emission}
Diffuse Galactic H$\alpha$ is thought to be a good tracer
of diffuse free-free emission
since both are emitted by the same ionized medium and 
both have intensities proportional to emission measure 
(the line of sight integral of the free electron density squared, 
$\propto \int N_e^2 dl$).

The intensity of H$\alpha$ emssion is given \cite{Reynolds},\cite{Leitch} by 
\begin{equation}
I_\alpha = 0.36 R ( {EM \over cm^{-6} pc} ) ({T \over 10^4~K})^{-\gamma}
\end{equation} 
for $T \leq 2.6 \times 10^4~K$
and where $\gamma$ varies from 0.9 for $T_e \leq 2.6 \times 10^4$~K
to 1.2 for $T_e > 2.6 \times 10^4$~K,
\begin{equation}
1 R \equiv 1 Rayliegh \equiv {10^6 \over 4 \pi} {\rm photons/(cm^2 ~s~ster)} 
= 2.41 \times 10^{-7} {\rm ergs/(cm^2 ~ s ~ ster)}
\end{equation}
at a wavelength $\lambda_{H\alpha} = 6563$~Angstroms.

Combining the free-free and H$\alpha$ equations one finds
a relation between the low-temperature H$\alpha$ intensity
and the free-free emission
\begin{eqnarray}
T_b^{ff} &=& 1.68 \mu K ~ < g_{ff} > ~ ({T \over 10^4~K})^{0.4~to~ 0.7} 
( {\lambda \over 1~cm})^2 ( { I_\alpha \over R } ) \\
&\approx & 7 \mu K ({T \over 10^4~K})^{0.55~to~0.85}
( {\lambda \over 1~cm})^{2.1} ( { I_\alpha \over R } )
\end{eqnarray}
so that for example
$ T_b^{ff}(30~GHz) = 7 \mu K ( { I_\alpha / R } ) $, 
$ T_b^{ff}(45~GHz) = 3 \mu K ( { I_\alpha / R } ) $, 
and $ T_b^{ff}(53~GHz) = 2 \mu K ( { I_\alpha / R } ) $.
A typical measured value for $I_\alpha$ is order of 1~$R$.

\subsection{Results from H$\alpha$ Surveys}

The major H$\alpha$ structures form the well-known Local (Gould Belt)
System which extends 30$^\circ$-40$^\circ$ from the plane at positive $b$
in the Galactic centre and at negative latitude in the anticentre.
The HI and dust in the Local System may be traced to 50$^\circ$
from the Galactic plane.
Other H$\alpha$ features are also found extending 15$^\circ$-20$^\circ$
from the plane \cite{Sivan}.

Quantitative measurements are now available from accurate spectroscopy
(e.g. Reynolds 1992, Bartlett et al. 1997).
To first order the H$\alpha$ may be approximately modelled as a layer parallel 
to the Galactic plane with a half-thickness intensity of 1.2 Rayleigh (R).
The rms variation in this H$\alpha$ emission is roughly 0.6R on degree scales.

The four-year COBE DMR sky maps at different frequencies have been
utilized to isolate emission with antenna temperature
which varies proportional to frequency to the -2.15 power
($\propto \nu^{-2.15}$) \cite{Kogut} 
in an attempt
to provide a large angular scale map of free-free emission at 53 GHz.
This low-signal to noise map is consistent with the H$\alpha$
large scale model with a free-free half height amplitude of $10 \pm 4\,\mu$K.
The rms free-free signal on a 7$^\circ$ scale was estimated to be
$\Delta T_{\rm ff} = 7 \pm 2 ~\mu$K.

The H$\alpha$ images of the NCP area made by Gaustad et al. (1996) have been
analyzed by Veeraraghavan \& Davies (1997) to provide an estimate
of the spatial power spectrum on scales of 10$^\prime$ to a few degrees.
The power law index is -2.3 $\pm$ 0.1 over this angular range.
The rms amplitude is 0.12 cosec($|b|$) Rayleighs on 10 arcmin scales.

It has been assumed that the free-free component could be modeled using 
measurements of H$\alpha$ emission, (e.g. \cite{Simonetti}, 
\cite{Reynolds92}) which measure the density of free electrons. 
However, recent results showing more significant correlation of dust and 
apparent free-free emission have been interpreted as presenting us 
with a surprize\cite{Kogut}\cite{Costa}\cite{Leitch}.
Alternative explanations have been suggested including
rotating dust grains\cite{Draine}.
It still appears that mapping the H$\alpha$ emission is the best way
to determine the free-free emission and separate out any other component.

Although Galactic H$\alpha$ emission is correlated with the
Galactic free-free emission,
it is not straightforward to estimate the Galactic free-free emission
from the H$\alpha$ sky maps.
The H$\alpha$ sky maps are contaminated with H$\alpha$
from the Earth's geocoronal emission.
The geocoronal emission varies both diurnally
and seasonally with the solar Lyman$\beta$ flux.
This variation ranges from 2 to 25 Rayleigh.

The geocoronal and Galactic H$\alpha$ emissions are separable
in principle, since the Doppler shift of the Earth's motion around the 
solar system separates the lines.
Making this separation requires spectral measurements
and is less effective as one moves toward the ecliptic poles
where the separation is negligible.
There is also the issue of another atmospheric line (OH) 
that partially blocks the positive Doppler shift side
so that only during restricted seasons can part of the geocoronal
effect be removed well in this manner.

\section{Galactic Corona and High Temperature Regions}
The ROSAT 1/4 keV survey of the soft x-ray background is being used to map
out the hot gas in the local bubble but more importantly
to show the existence of the long hypothesized \cite{Spitzer56} 
hot Galactic corona.

In 1956 Spitzer\cite{Spitzer56} suggested that the Galactic halo 
was filled with hot gas.
A primary argument was the existence of cool gas clouds high above 
the Galactic plane, which he reasoned must be confined by
a hotter ambient medium.
This hot medium must be continually being resupplied with energy
as cooling would soon set in.
The favored current model is of fountains of hot ionized material
produced by large groups of supernovae.

The dust in the Draco complex appears as a ``shadow'' in the
soft x-ray background providing direct evidence that a large
portion of the soft x-ray background comes from the Galactic halo.
The ROSAT soft x-ray data and HI maps have been used to model
the hot ionized interstellar medium, e.g. \cite{Sider},
to obstain estimates of the temperature and emission measure ($EM$).

The x-ray temperature estimates are on the order of $10^{6.2\pm0.2}$~K
which are consistent with the predictions based upon 
the virial temperature $T_{virial} \sim 2 \times 10^6$~K of the halo.

\section{Emission form the Magellanic Stream}

The Magellanic Stream is a long filament of H I clouds which stretches
over 100$^\circ$ across the sky which trails behind the Magellanic Clouds
in their orbit around the Galaxy\cite{Mathewson}.
In ram pressure explanations for the origin of the Magellanic Stream, 
the Stream is swept out of the Magellanic Clouds by the diffuse ionized
corona of the Galactic halo. 
The Stream is a chain of clouds connected by lower-density gas.
These clouds generally have a high-density concentration and gradient
on the leading edge\cite{Jones} 
and H$\alpha$ emission is observed on some of these clouds\cite{Weiner}.
The H$\alpha$ emission is best explained by ram pressure heating from the 
hot Galactic corona.

\section{Estimates of Free-Free Emission}

Table \ref{Tabest} presents measurements and estimates 
of the Galactic ionized emission regions.

\begin{table}
\caption{\label{Tabest} Estimates of H$\alpha$ \& Free-free emission}
\begin{center}
\begin{tabular}{l r c c c}
\hline
Region & T & $ EM $ & $I_\alpha$ & $T_b^{ff}$(30~GHz) \\
       &(K)&($cm^{-6} pc$) & ($R$) & ($\mu K$)   \\
\hline
Local Bubble& $10^{5.9}$ & 0.004 \cite{Sider} & $3 \times 10^{-5}$ & $10^{-3}$ \\
Disk Region ($\sim$ 1 kpc) & $10^4$ & 2.9 csc$|b|$ \cite{Reynolds}& 1  &  5 - 40 \\
Halo & $10^{6.2}$ & 0.024 \cite{Sider} & $9 \times 10^{-5}$ & $7 \times 10^{-3}$ \\
Magellanic Stream & $10^4$ & 0.5-1.0 \cite{Weiner} & 0.2-0.37$\pm$0.02 & 2-4 \\
Local Group Corona& $\sim 10^6$ &  & & $\sim 0.1$ \\
\hline
\end{tabular}
\end{center}
\end{table}
Using the formulae relating the free-free emission to
the plasma temperature and emission measure
and to H$\alpha$ emission one derives the signals shown.
The numbers for the Galactic free-free emission
derived this way are a factor of two below 
what one finds using the slope (fitting to csc$|b|$) COBE DMR free-free map.
The DMR free-free map does not have a strong signal-to-noise ratio
and shows a high degree of correlation 
with the Galactic dust emission\cite{Kogut}
which may or may not be free-free emission\cite{Draine}.
For that reason it is more reliable and consistent 
to trace the properties of the ionized interstellar medium
through its H$\alpha$ emission.
The major question is: How much hot ($T >> 10^4$~K) gas is there
in the interstellar medium? Could there be enough to make up 
the factor of two? 
The x-ray and pulsar dispersion measurements
put a tight limit on the possible additional free-free emission.
The pulsar dispersion measurements are consistent with an emission
measure of approximately 1 csc$|b|$ /($cm^{-6}pc$) if all the ionization
is spread evenly and uniformly. 
The estimate is higher by the fractional filling factor.
The filling factor is estimated to be of order 0.1 to 0.4, 
increasing with increasing Galactic latitude.

\begin{figure}[h]
\epsfxsize=6.0truein
\epsfbox{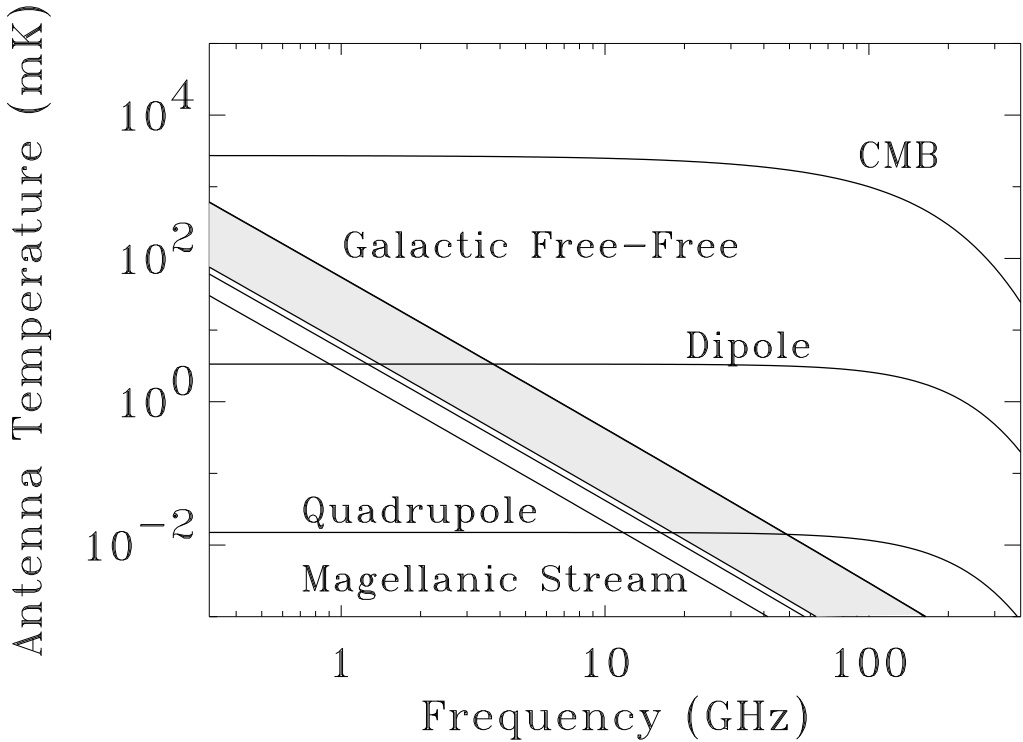}
\caption{
Graph showing frequency dependence and estimated signal levels
of Galactic free-free emssion for medium and high lattitudes
as a filled band and an open band for the Magellanic Stream clouds
as well as that of the cosmic microwave background and its features.
}
\label{FigFF}
\end{figure}

\section{X-ray Emission Information}
The volume emission by thermal bremsstrahlung is
\begin{eqnarray}
{d E \over dt d \nu d V}
&=& [ {32 \pi e^6 \over 3 m c^3 } ] ({2 \pi \over 3 k m_e})^{1/2}
T_e^{-1/2} Z^2 e^{-h \nu /k T_e} < g_{ff}> \\
&=& 6.8 \times 10^{-38} ~ Z^2 N_e N_i T_e^{-1/2} e^{-h \nu /k T_e} < g_{ff}> 
{\rm erg s^{-1} cm^{-3}}
\end{eqnarray}

Integrating over the line of sight one finds
\begin{eqnarray}
{d E \over dt d\nu d A} 
&=& 6.8 \times 10^{-38} ~ \int N_e N_i dl ~ T_e^{-1/2} Z^2 e^{-h \nu /k T_e} 
< g_{ff}>  {\rm erg s^{-1} cm^{-3}} \\
&=& 2.12 \times 10^{-20} ~ {EM \over cm^{-6} pc } 
T_e^{-1/2} Z^2 e^{-h \nu /k T_e} < g_{ff}> {\rm erg s^{-1} cm^{-2}} \\
&=&  3.3 \times 10^3 {EM \over cm^{-6} pc } Z^2 T_e^{-1/2} e^{-h \nu /k T_e} 
< g_{ff}> {\rm keV s^{-1} cm^{-2} keV^{-1}}
\end{eqnarray}

X-ray emission is a good tracer of radio free-free emission when the
plasma is sufficiently hot to produce x-rays via thermal bremstrahlung.
That is they both arise from the same mechanism but are the opposite
extremes of $h \nu / k T_e$. 
Thus they have essentially the same coefficients except 
the Boltzmann suppression factor is very important for x-rays.
One can then use the x-ray observations to provide an estimate
and upper limit versus temperature for the radio free-free emission.

\begin{figure}[h]
\epsfxsize=7.0truein
\epsfbox{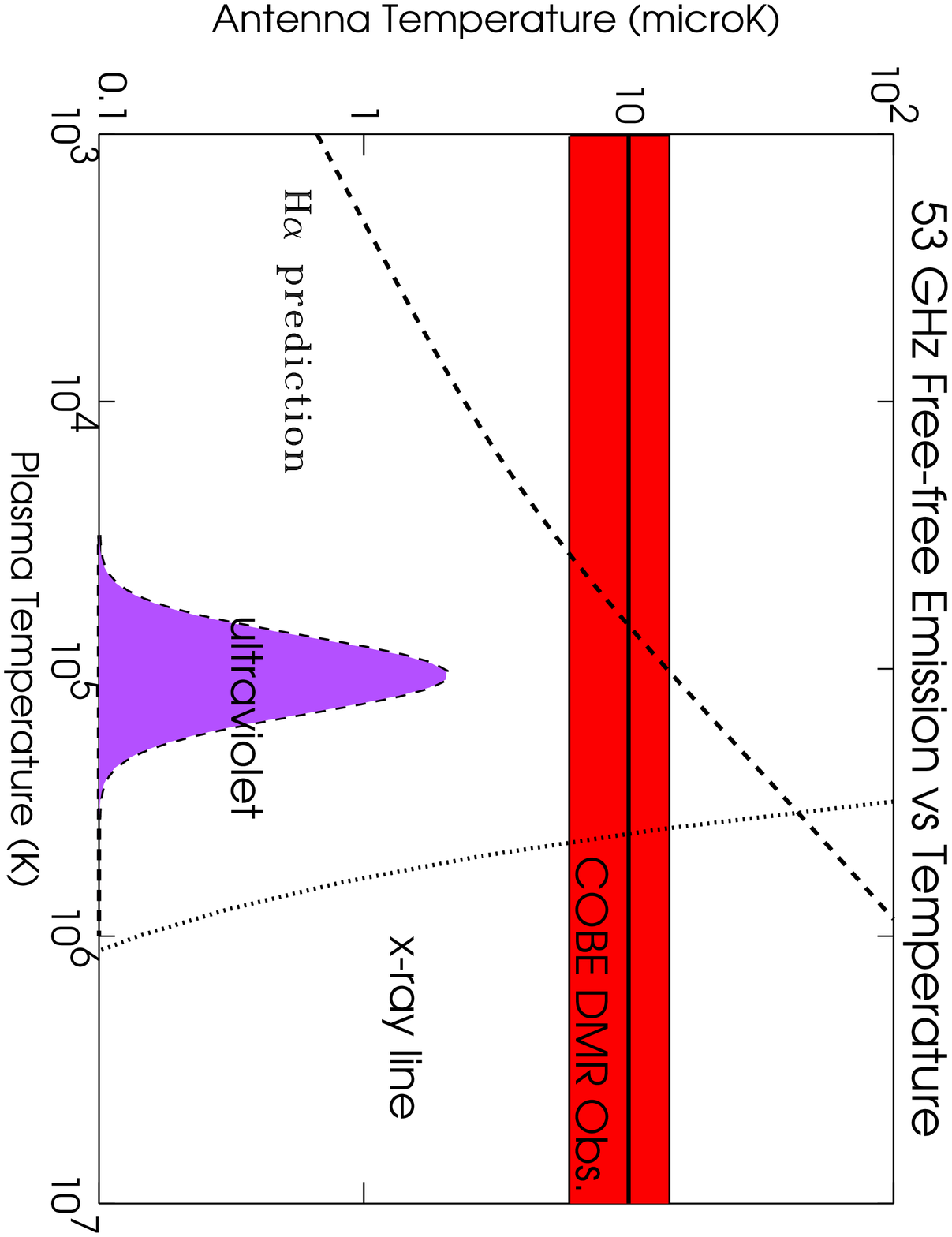}
\caption{
Graph showing plasma temperature dependence of the predicted 53 GHz
free-free emission antenna temperature for the mean observed
H$\alpha$ emission (1.2 R).
Also shown is the limits derived from x-ray and extreme ultraviolet emission
and the fit to the COBE DMR observations.
}
\label{FigFFX}
\end{figure}

A summary of the observations, H$\alpha$ prediction, and the x-ray limits
are shown in Figure \ref{FigFFX}.
(Note that the OVRO \cite{Leitch} observations of about 200 $\mu$K 
at 14.5 GHz extrapolate to be roughly 13 $\mu$K at 53 GHz;
however, since the signal is at such a small angular scale,
one would have predicted a smaller number by a factor of roughly 10.)

\section{Ultraviolet Observations}
Current H$\alpha$ and x-ray observations leave room for plasma 
in the $T_e \sim 10^5$~K range to produce significant radio free-free.
One would not be surprised that a rising Galactic fountain of hot
supernova gas would cool from $10^6$~K to this level.
However, measurements in the ultraviolet can constrain this possibility.
Ultraviolet observations of O {\sc VI} \cite{Hurwitz}, \cite{Dixon}
indicate the presence of halo hot gas in the temperature range
$ 5.3 \leq log(T_e) \leq 5.8 $.
The mean density of the gas is of order 0.01 cm$^{-3}$ with scale height
of less than 1 kpc and filling factor greater than 0.1.
This can be used to limit the radio free-free emission.
This limit is generally below 0.1 $\mu$K.
The plot in Figure \ref{FigFFX} shows a significantly higher limit
both to be very conservative and so that the limit would show on the plot.

\section{Geocoronal Free-free \& H$\alpha$ Emission}

The Earth's corona produces H$\alpha$ emission.
About 12\%\ of the hydrogen atoms excited by solar Lyman $\beta$ photons 
return to their ground state through the emission 
of H$\alpha$ photons \cite{Shih}.
This geocoronal emission varies both diurnally and seasonally
with the solar Lyman $\beta$ flux.
The amplitude variation ranges from about 2 to 25 Rayleighs.

Geocoronal H$\alpha$ emission is not expected to be a tracer
of geocoronal free-free emission as most of the hydrogen is not 
significantly ionized but excited via solar Lyman$\beta$.
The mean temperature of the geocoronal hydrogen ranges from 900 K to 1300 K.

The geocoronal H$\alpha$ emission is a potential interference 
for observing the Galactic H$\alpha$ emission.
Doppler effect allows separation because of the motion of the Earth 
around the Sun and the relative motion of the plasma.
However, such a separation requires high resolution spectral measurement
and analysis instead of simple imaging.

\section{H$\alpha$ \& Free-Free vs. Dust Emission Correlation}
A number of groups have found a significant correlation between dust
and microwave ``free-free'' (antenna temperature spectral index $\sim -2$)
emission by cross correlation between 100~$\mu$m IRAS and DIRBE maps
and the observed emission 
\cite{Kogut},\cite{Leitch},\cite{Costa},\cite{Veeraraghavan}.
and H$\alpha$-dust correlation \cite{Koguta}, \cite{McCullough}.
Table \ref{Tabcor} provides a summary of results on the dust vs H$\alpha$ \& 
free-free emissions.
Table \ref{Tabcor} shows clearly that the estimated ``free-free'' emission 
correlation to the dust implies a larger total signal than the H$\alpha$.
This is the same conclusion one tentatively reaches from Figure \ref{FigFFX}
which gives the predicted signal level or upper limits from the 
H$\alpha$, ultraviolet, and x-ray observations.

\begin{table}
\caption{\label{Tabcor} Correlation of Dust vs Free-free \& H$\alpha$ Emission}
\begin{center}
\begin{tabular}{l c c c c}
\hline
\multicolumn{5}{c}{\bf Free-free--Dust Correlation} \\
Authors & $T_{ff}/I_{100 \mu m}$ & frequency $\ell$& $b$ \\
        &( $\mu$K (MJy/sr)$^{-1}$)        & (GHz) & (deg) & (deg) \\
\hline
Leitch et al. \cite{Leitch} & 75 & $(\nu/23)^{-2.2}$ & & NCP \\
Kogut et al. \cite{Kogut} & $18.06 \pm 2.54$ & 31.5 & & $|b| > 20^\circ$  \\
                          & $ 6.88 \pm 1.40$ & 53 & & \\
                          & $ 2.76 \pm 1.61$ & 90 & & \\
Costa et al. \cite{Costa} & $ 15.0 \pm 8.1$  & 40 & & NCP \\
\hline
\multicolumn{5}{c}{\bf H$\alpha$--Dust Correlation} \\
Authors & $I_{H\alpha} /I_{100 \mu m} $& frequency & $\ell$& $b$ \\
        &( Rayleigh (MJy/sr)$^{-1}$) & & (deg) & (deg) \\
Kogut 1997 \cite{Koguta}  & $0.85 \pm 0.44$ & H$\alpha$ & $144^\circ$ & $-21^\circ$ \\
Kogut 1997 \cite{Koguta}  & $0.34 \pm 0.33$ & H$\alpha$ &  & NCP \\
McCullough \cite{McCullough} & $ 0.79^{+0.44}_{-0.15}$ & H$\alpha$ & $71^\circ$
& $-67^\circ$ \\
\multicolumn{5}{c}{\bf Implied from Free-free $T_e \approx 10^4$~K} \\
Leitch et al. \cite{Leitch} & 6 & & & NCP \\
Kogut et al. \cite{Kogut} & $ 2.0 \pm 0.5 $ & 50 & & $|b| > 20^\circ$  \\
Costa et al. \cite{Costa} & $ 3.6 \pm 2   $ &    & & NCP \\
\hline
\end{tabular}
\end{center}
\end{table}

There is also an inconsistency between the estimates from the microwave 
estimates of ``free-free'' signal emission unless there is significantly more
correlation between the ``free-free'' signal and the dust emission 
at smaller angular scale.

\section{Conclusions}
The estimation of radio (e.g. 53 GHz) free-free emission
is not a completely settled issue.
It is possible that some of the signal seen is due to rotating dust
or some other no yet understood source.
However, it is possible that the signal level and the correlation
with dust will be readily accounted for by simple
emission from warm ($T_e \sim 10^4$~K) plasma
which is well traced by its H$\alpha$ emission.
The issue will be resolved when there are both
high quality H$\alpha$ and microwave observations
and the results can be carefully cross correlated.
At that point we will be able to determine
what is the level of the other sources.

{\bf Acknowledgments.} --- This work supported in part by the DOE
contract No. DE-AC03-76SF00098 
through the Lawrence Berkeley National Laboratory
and NASA Long Term Space Astrophysics Grant No. 014-97ltsa.


\begin{thebibliography}{99}
\bibitem{Banday} A. Banday, and A. Wolfendale, 1990 MNRAS, 245, 182
\bibitem{Bartlett} Bartlett, J., Marcelin, M, et al. Strasbourg-Marseille Collaboration
\bibitem{Costa} A. de Oliveria-Costa et al. 1997 ApJ, 482, L17, astro-ph/9705090
\bibitem{Dixon} Dixon, W. Van Dyke \& Davidsen, A.F., 1996 ApJ, 465, 288.
\bibitem{Draine} B.T. Draine \& A. Lazarian 1997 ApJ astro-ph/9710152
\bibitem{Jones} Jones, B.F., Klemola, A.R., \& Lin, D.N.C., 1994 A.J. 107, 1333.
\bibitem{Gaustad} Gaustad, J., McCullough, P. \& van Buren, D., 1996. PASP,
108, 351.
\bibitem{Hurwitz} Hurwitz, M. \& Bowyer, S. 1996 ApJ, 465, 296. 
\bibitem{Kogut} A. Kogut et al. 1996 ApJ, 460, 1, astro-ph/9509151
\bibitem{Koguta} A. Kogut 1997 AJ, 114, 1127 astro-ph/9706282
\bibitem{Leitch} E. Leitch, A.C.S. Readhead, T.J. Pearson, \& S.T. Myers 1997
astro-ph/9705241
\bibitem{Mathewson} Mathewson, D.S. Cleary, M.N., \& Murray, J.D., 1974, 
Ap. J. 190, 291.  
\bibitem{McCullough} McCullough, P.R. 1997 A.J., 113, 2186 astro-ph/9703128
\bibitem{McKee} McKee, C.F. 1993 `Back to the Galaxy,' ed. S.S. Holtz \& 
F. Verter (New York) AIP 499   
\bibitem{Oster} Oster, L. (1961) Rev. Mod. Phys. 33, 525
\bibitem{Reynolds} Reynolds R.J., et al. 1998, WHAM (Wisconsin H$\alpha$ Mapper)
home page: http://www.astro.wisc.edu/wham/
\bibitem{Reynolds92a} Reynolds R.J., 1992, ApJ, 392, L35.
\bibitem{Reynolds92}R.J. Reynolds, and D.P. Cox, 1992, ApJ.L, 400, 33
\bibitem{Rohlfs} Rohlfs, K. (1990) ``Tools of Radio Astronomy'' Springer-Verlag
\bibitem{Rybicki} G.B. Rybicki, and A.P. Lightman, 1979, 
'Radiative processes in astrophysics', (Wiley, New York USA)
\bibitem{Sider} S.D. Sider, T.J. Sumner, J.J. Quenby, \& M. Gambhir 1996,
Astron, Astrophys 305, 308-315.
\bibitem{Simonetti} J.H. Simonetti, et al., 1996, ApJ.L., 458, 1
\bibitem{Sivan} Sivan, J.P., 1974. AA Suppl. 16, 163.
\bibitem{Shih} Shih, P., Roesler, F. \& Scherb, F. 1985 J. Geophys. Res. 90, 477.
\bibitem{Spitzer56} Spitzer, L. 1956 Ap. J. 124, 20.
\bibitem{Spitzer78} Spitzer, L. 1978 ``Physical Properties of the Interstellar
Medium" (New York: John Wiley \& Sons)
\bibitem{Spitzer90} Spitzer, L. 1990 ARA\&A 28, 71 
\bibitem{Valls-Gabaud} Valls-Gabaud, D. 1998 Pub. Astron. Soc. Australia, 
Vol. 15 
\bibitem{Veeraraghavan} Veeraraghavan, S. \& Davies, R.D., 1997 in Particle
  Physics and the Early Universe, eds. Bately, R., Jones, M.E. \& Green, D.A.
\bibitem{Weiner} B.J. Weiner \& T.B. Williams 1997, Astron J., astro-ph/9512017

\end{thebibliography}
\end{document}